# On the orientation of Roman towns in Italy

**Giulio Magli**
**Dipartimento di Matematica del Politecnico di Milano**
**P.le Leonardo da Vinci 32, 20133 Milano, Italy.**

**As is well known, several Roman sources report on the existence of a town foundation ritual, inherited from the Etruscans, which allegedly included astronomical references. However, the possible existence of astronomical orientations in the layout of Roman towns has never been tackled in a systematic way. As a first step in this direction, the orientation of virtually all Roman towns in Italy (38 cities) is studied here. Non-random orientation patterns emerge from these data, aiming at further research in this field.**

**1. The planning of a Roman town**

During the republican period and the first imperial period (roughly from the V century b.C. to the I century a.D.) the Romans founded many towns, or *colonies.* The foundation of a town was an efficient method in order to establish permanent control on a newly conquered land, and a consistent amount of people (up to 5000 families) was assigned to the new towns in accordance with a legal act called *deductio.* The urban plan of the settlements was always the same: the layouts of the newly founded towns were indeed planned in accordance with the so called *castrum* (i.e. military camp) structure.[1] A colony could also be deducted "on" a pre-existing, non-Roman town, either conquered or allied; also in this case however, the layout was usually re-designed in accordance with the Roman standards (see e.g. Sommella 1988).

The city boundaries of a *castrum* town formed a rectangle, usually not very stretched in one direction or the other, so that we can refer generally to this kind of town planning as to a *squared* one. The rectangle was delimited by the walls and the internal streets were organized in a orthogonal grid forming the inhabited quarters (*insulae*). As is well known, orthogonal town planning – which has been a common feature of many cultures, for instance the Hindu one - was theorized in the classical world by the Greek architect Hippodamus of Miletus around the half of the V century, and applied strictly in all Greek colonies (Castagnoli 1971). Beyond any doubts it was applied by the Etruscans as well, as it is shown by the excavations at Misa, the unique known Etruscan town which was conquered by the Celts and abandoned before the Roman expansion (Mansuelli 1965).

Some authors believe that the Etruscans simply copied the orthogonal plan from the Greeks, but in any case it is reasonable to think that the Romans inherited the orthogonal layout from the Etruscans, rather than from the Greeks. In addition to the orthogonal layout however, the inspiring principle of the Roman *castrum* was based on the existence of two main orthogonal roads, called *cardus* and *decumanus.* Thus, in Roman towns, the orthogonal layout was actually *quadripartite*: at the end of the four main roads, four main gates were located, while the centres of the social and religious life (the Forum and the main temple or Capitolium, respectively) were preferably placed at (or near) the intersection of the main roads (see e.g. Rykwert 1999). To fix ideas about a Roman castrum town, we can look at Fig. 1, where, as an example, the plan of Augusta Pretoria (today's Aosta, founded around 25 b.C.) is shown.

---

[1] This terminology is slightly misleading since, as noticed by Castagnoli (1971), it was probably the layout of the towns to inspire that of the camps, and not vice-versa.

## 2. Astronomical references in written sources and in landscape planning

The foundation of a new town followed a ritual, which has been described by many Roman writers, like e.g. Ovid and Plutarch. All these authors refer themselves to the foundation of Rome, so that in a sense the foundation of a Roman town might be seen a sort of *replica* of a primeval foundation. This ritual, as is universally known, comprised the observation of the flight of the birds and the tracing of the boundaries by ploughing a furrow. The art of taking auspices from the flight of the birds was ruled by the *Etrusca Disciplina*, the collection of writings of the Etruscan religion, which was thought of as having being revealed to humanity by the gods (the books are long lost, but accounts on them have survived, for instance in the work *De Divinatione* by Cicero). A fundamental part of all the rituals of the aruspexes was the individuation of the *auguraculum*, a sort of terrestrial image of the heavens (*templum*) in which the gods were "ordered" and "oriented" starting from north in the hourly direction (an intact example of *auguraculum* has been found in the city of Bantia, see Torelli 1966). The individuation of the templum thus required astronomical orientation to the cardinal points (Aveni & Romano 1995, Pallottino 1997); at the corresponding "centre" (*mundus*) a deposit of foundation containing first produces of the fields and/or samples of soil from the native place of the founders was buried. Archaeological proofs of such foundation rituals have never been found in Rome, where several subsequent rearrangements have probably cancelled the traces of the original *Roma Quadrata* on the Palatino hill (but see Carandini 1997); however, proofs of foundation deposits have been discovered in the excavations of the Etruscan towns Misa (Mansuelli 1965) and Tarquinia (Bonghi-Jovino 2000), while for the Roman period clear traces of the foundation ritual and of his connections with astronomy have been found in Cosa and in Alatri (Brown 1980, Aveni & Capone 1985, Magli 2006, 2007).

Explicit references to the Etruscan ritual are also present in the so called *Corpus Agrimensores*, the collection of technical treatises on the procedures of landscape planning and division. These procedures were called *centuriation* and consisted in a rigorous geometric division of the lands in squared lots of one *actus* (710 meters) each side (in many places, the "fossil" traces of this geometric division are still visible today, trough aerial photographs and/or present-day cadastral maps). Centuriation is very old: the first known to us is that of Tarracina (today Terracina) which goes back as far as the middle of the IV century BC (Castagnoli 1971). The dimensions of a centuriation were sometimes really impressive, if not nearly incredible, such as, for instance, those of the land division of Tunisia (254 x 110 Kms) made by the Agrimensores of the third *Legio Augusta* in the first half of the first century BC (Decramer & Hilton 1998, Decramer et al. 2001). Exactly as in the case of the towns, the orthogonal grid was based on two main roads; their intersection was the centre of a coordinate system in which each point was individuated by the progressive numbers of the intersecting lines.

According to what the Agrimensores say in their treatises, the tracing of the two main roads, and consequently the orientation of the whole grid, could occur in accordance with several different criteria, and different orientations are actually found in researches on ancient topography. For instance, a grid having one main axis disposed along a prominent landscape feature was sometimes needed. A good example is the centuriation of Luni (near the coast in Liguria, Italy). The available land was comprised in a strip between the sea and the mountains; thus, one of the main axis had to be parallel to the average direction of the coast, otherwise an exceeding number of incomplete lots would have come out. A grid could also be disposed by putting one of the main axes along a pre-existing preferred direction, typically that of a main connecting road and/or one of the two main roads of a town. Also in this case, indeed, the use of the existing road as one of the two directions, or even as one of the two main roads, was practically obliged, otherwise the lots would have been crossed by the road/town's boundaries. Several examples of this kind of orientation can be seen along the Via Appia, the road connecting Rome with Capua built around 312 b.C.

Interestingly enough, in their texts the Agrimensores look somewhat "unhappy" of making centuriation in accordance of such practical criteria. Indeed, they all explicitly, repeatedly and proudly state that their art was inherited from the rituals of the Etruscan Disciplina, and that it was thus founded on a *symbolic* conception of the human landscape; perhaps stretching things a little far, one could say that the process of centuriation was connected, at least in their mind, with an idea of *sacred space*. For instance, Frontinus states

Limitum prima origo, sicut Varro descripsit, ad disciplinam Etruscam; quod aruspices orbem terrarum in duas partes diviserunt… altera[m] linea[m] a septentrione ad meridianum diviserunt terram…. Ab hoc fundamento maiores nostri in agrorum mensura videntur constituisse rationem. Primum duo limites duxerunt; unum ab oriente in occasum, quem vocaverunt decimanum; alterum a meridiano ad septentrionem, quem cardinem appellaverunt.

Which means (literal translation by the author):

The first origin of the art of tracing limits, as stated by Varron, comes from the Etruscan *Disciplina*, where the aruspexes divide the earth into two parts…and [in two other parts] with another line from the north along the meridian. From this foundations our predecessors took the art of measuring the lands. First, two streets will be traced: one from the east, which will be called decumanus, the other on the meridian from the north, which will be called cardus.

Thus, according to the Agrimensores, their discipline included a symbolism connected with the sky, and this symbolism was ancient as much as the rules of the *Disciplina*. It goes without saying that indeed many examples of centuriations oriented to the cardinal points do exist: for instance, the centuriation of *Augusta Raurica* (Swiss), dated to the first decades of our era, and those of Capua (II century b.C) and Nola (beginning I century b.C.) in southern Italy. Further to this, orientation of the decumanus to one of the two solstices azimuth of the sun is documented as well, for instance near the city of *Cartago* (the Roman colony at the site of Cartage).
In spite of this quantity of instances, the existence of astronomical references in the planning of Roman towns has been repetitively negated, or admitted only for functional, rather than symbolic, motivations (see e.g. the reference book by Adam 1999). This position assessed after the work by Le Gall (1975), who maintained that:

1) the Agrimensores just *invented* the symbolic and sacred content of their science, claiming for a derivation from the Etruscan Disciplina
2) the astronomical orientation mentioned by them regards in any case only the centuriation procedure, and therefore cannot be extended to the towns
3) as a consequence, there is no astronomical content in the planning of the roman towns

To sustain his theses Le Gall presented a list of data scattered from York, Brittany (54 degrees of latitude north) to Cuicul, Algeria (36 degrees). However, as we shall see later, data having a wide spread in latitude make it difficult to extract significant information on solar orientations. Further to this, Le Gall data are extremely few and very heterogeneous: two centuriations, ten towns and two military camps. It is therefore clear, at least in the opinion of who writes, that this problem fully merits a full, complete reassessment and analysis. The present paper is actually only a first step toward this analysis, because it concentrates only on Italy. However, I do believe that the evidences I am going to present are at least sufficient to show that astronomical orientation has been in many cases a fundamental component in the planning of Roman towns.

**3. The orientation of Roman towns in Italy**

I have collected (Table 1) the following data:

1) The orientation of the orthogonal grid of the towns which are certainly of Roman foundation in Italy. I do *not* take in consideration all the Roman towns, but only those (actually, nearly all) in which at least the orthogonal grid, if not the two main roads, is still clearly discernible, and I hope to have considered *all* of the latter (I would, of course, be grateful of the advice on any town eventually missing).
2) The orientation of the orthogonal grid of all the towns which are not (or not with certainty) of Roman foundation in Italy, but where a orthogonal layout was certainly super-imposed by the Romans.

In both cases, by "orientation" I mean the angle formed with due east by that axis which locates within the squared angle bisected by due east, counted positively in the anti-hourly (northerly) direction . This is not the standard measure of azimuths (customarily measured positively from north to east) but, as we shall see in a while, in this way the data become more easily legible. Most of orientations have been extracted from available archaeological maps; in some cases I have controlled their value on site using a precision magnetic compass (accounting of course for magnetic deviation, but this is nearly zero at the time of writing in Italy) and I actually found them reasonably precise (within one degree of difference). Further, for a few of these towns archaeoastronomical research is already available, and in these cases the data are usually collected with a teodolite and are consequently much more precise than the average precision which has to be generally expected here (not better than plus or minus one degree). These data are marked with a "Y" in the last column of Table 1, and the corresponding references are cited in the text below.
To explain the way in which the data have been arranged, a few technical remarks are necessary (I apologize for the details which I am going to discuss, since most of the readers would consider them as childish considerations; however I would like to make the paper readable also to readers who are not familiar with solar astronomy).
We are interested here in the yearly movement of the rising sun at the eastern horizon during the course of the year. I will completely avoid any consideration which would actually alter the "idealized" view I am going to adopt, thus, in particular, I will not take into account atmospheric refraction, and the possible existence of a local non-flat horizon. The reason is that this paper of course does not aim to a complete archaeoastronomical evaluation of each site, a task which would require a huge field-work campaign. My aim here is rather to investigate on the existence or not of an "archaeoastronomical phenomenon" which, once proved, would certainly show the need for a more complete analysis. Thus, I am going to consider the movement of the rising sun as the idealized movement of a point in a symmetric arc-sector having due east at the centre. It is important to remember that the movement of this point does not occur with a constant velocity: it is very slow near the solstices, and very fast near the equinoxes; with our choice for the measures of angles, the arc-sector spanned in the course of one year is individuated by an angle $\alpha$ and its opposite $-\alpha$; the value of $\alpha$ of course increases with the latitude of the site (the minimum, at the equator, is around 26 degrees). Consider now two orthogonal axes – a cross - having the same centre of the cardinal axes but rotated with respect to them, and search for an angle such that the sun will not raise or set in alignment with any of the two axis of the cross, never, during the course of the whole year. It is easily seen from Fig. 2 that the answer to this question is as follows. It is *possible* to find orientations of this kind, if and only if the latitude of the site is such that $\alpha$ is less than 45 degrees; in this case, one of the axis of the cross must lie within the shaded sector of amplitude $\beta=90-2\alpha$. It goes without saying, therefore, that any castrum town constructed at a latitude such that $\alpha$ is near to, or greater than, 45 degrees, is always oriented to the rising sun in some day of the year (this latitude is around 55 degrees). Further, it is certainly preferable to compare data which come from sites whose latitude is not too different, so that their reference value $\alpha$ is comparable. In particular, if we want to understand - as a preliminary step – if solar orientation actually *was* there, it is better to work in a strip of latitudes where a sufficient amount of data is available and the angle $\alpha$ is reasonably less than 45 degrees (for the data considered in the present paper, the value of a varies within a interval of around 5 degrees from 35 to 30).

For each town the following are reported:

1) Latitude
2) Latin name
3) Today's name, and approximate date of foundation or Roman re-foundation
4) Orientation of that axis which lies within 45 degrees from due east, indicated with NE (north of east) or SE (south of east) respectively
5) Evaluated amplitude at the solstices at the date of foundation. This amplitude is reported only for those towns in which one of the axis locates within ten degrees from it.
6) Existence of already published archaeoastronomical research on the town orientation, indicated with a Y; the corresponding references are cited in the text below

The following observations can be easily made:

1) There are only three northern orientations out of the solar sector. These are today's Pesaro, Rimini and Senigallia, towns which lie relatively close, on the west coast of Central Italy. It might be that their orientation was dictated by similar, and special, considerations, which anyway await for further investigations.
2) Only two towns lie near the summer solstice sunrise line. These are Verona and Vicenza for which deliberate solar orientation has been actually already proved (Romano 1992); also these two towns are geographically close and where founded in the same period; also in this case, therefore, special considerations – not known at present - should apply
3) There is a family of 14 towns located near the cardinal points, in a sector of around 10 degrees of amplitude on both sides of due east. Of these, 9 lie in the south of east sector.
4) There is a family of 9 towns which locates near the winter solstice sunrise line. Of course the rising sun follows a southerly oriented path in the northern hemisphere, and therefore all such towns, oriented near the solstice line or with slightly *greater* azimuth, are oriented to the "climbing" sun, at a relative low altitude. Among these towns, two (Bene Vagenna and Norba) have been actually already shown to be aligned to winter solstice sunrise (Barale et al 2002, Magli 2006)
5) Thee is a vacancy of data in the intermediate sectors, in particular, no towns lie in the 19-29 south of east sector.

Mathematically, we can convince ourselves of the existence of a non-random distribution in presence of such a small sample of data inspecting how much the data differ from a binomial distribution (other tests, like the chi-square, require more populated samples).
First of all, we inspect the region around the cardinal points calculating the number of towns which should fall in a sector of 20 degrees of wideness. We thus have (38*20/90) ~ 8.4 towns expected, with a standard deviation around 2.5. A distribution can be considered to be statistically significant (i.e. non casual) from values of the order of the average plus or minus two standard deviations, and therefore, in this case, statistical significance starts at ~ 13.4 towns, whence this sector has 14. Orientation to cardinal points becomes more clear if only the sector south of east is considered. For a 10 degrees sector, we have (38*10/90) ~ 4.2 towns expected with a standard deviation of ~ 1.9. Statistical significance therefore starts around 8 towns, and the sector under exam has 9; in other words, it is actually this contribution which makes the region around due east statistically relevant. Significance also holds for the 30-39 degrees south of east sector cited above, which comprises nine towns as well.
The absence of towns in the sector between 19 and 29 degrees south of east is, of course, significant too (the average *minus* two standard deviations gives zero). Interestingly enough, it may be noticed that the solar dates corresponding to these azimuths locate in two periods which do not contain any relevant festivity of the Roman calendar, namely (very roughly, because the effective dates depend on the specific orientation and latitude) the second half of November and the second half of January. This observation can be compared with the fact that, instead, dates falling between 10 and 19 degrees north or south of east may indicate important festivals of the Roman Calendar. In particular, in the second half of February (orientations south of east) many important festivals took place: the *Parentalia* (Festival of the deads), the *Lupercalia* (Festival of Faunus), and the *Terminalia*, festival of the god Terminus, protector of the boundaries and of the city walls. It has been actually already proposed that the orientation of Bonomia (Bologna) was chosen in such a way that the sun was rising in alignment with the decumanus of the city on the day of *Terminalia* (Incerti 1999), and a fieldwork may lead to similar conclusions for other towns of this group as well. On the northern side there are of course too few data to draw conclusions; however, the distribution between 9 and 25 degrees NE is at least intriguing: only five towns, concentrated in only two angles. The corresponding dates fall into the period 10-30 of April which, of course, includes the foundation of Rome (21 April). From the archaeoastronomical point of view, this is the unique date for which an indubitable astronomical "alignment" is documented among the Romans: it is the ierophany which takes place in the Pantheon (see e.g. Belmonte & Hoskin 2002).

Some further comments on these data are in order.

1) The presence of non-random distribution shows that an hypothesis which has been put forward in the past, namely that the orientation of Roman towns could have been done towards the rising sun at the date of birth of the founder, has no possible validity (this is actually an idea which has nothing to do with the Roman mentality and was already rejected by Le Gall).

2) One may ask for the techniques which were effectively used by the surveyors in tracing the layout. A very simple and natural way of working with a fixed angle in a wide dockyard consists in choosing angles which have a rational tangent (that is, correspond to a rational ratio between the two legs of a right triangle). Indeed, in this way different persons working together do not need samples of the angle (a procedure which would result in propagating errors): it is sufficient for them to know the chosen ratio, and to trace a rectangle using an arbitrary unit of measure, to conform to the desired proportions. This method was, as is well known, used by the Egyptians (for instance, the slope of the Cheop's pyramid corresponds to 14/11) and there are proofs that it was used by the Agrimensores for landscape planning (see e.g. Cataldi 2004, Peterson 1992); perhaps proofs can be found also for the town's layout. It should be noted that in such a case astronomical orientation to cardinal points would become fundamental independently from the chosen orientation of the grid. In fact, to be sure to obtain the desired azimuth (at fixed amplitude of angle) in tracing the legs, the surveyors had to construct a "local point grid" oriented on the cardinal directions

3) Perhaps some angles with rational tangent may have been preferred to others for symbolic reasons, for instance those corresponding to Pythagorean triangles. This idea has already been proposed for the camps (Richardson 2005) and certainly deserves further studies, since much more accurate measures would be required as well as wider samples of data.

4) As far as the techniques for finding the meridian are concerned, it looks quite probable that these were solar-based and made use of the bisection of the angle formed by the shadows of a post (gnomon) at symmetric hours of the day. Indeed, techniques based on the stars[2] are usually much more precise than the average precision observed in the Roman world.

---

[2] There are essentially two possibile methods based on stars: observation of upper and lower culmination of circumpolar stars, and bisection of angles of rise and set of non-circumpolar stars (due to precession, there was no ``pole star`` - i.e. a star sufficiently close to the north celestial pole - in Roman times).

5) A final comment is in order about the possibility of lunar orientations, although this kind of orientation is never documented in the existing texts. The moon has four stations, analogue to the solar solstices, which are reached once every 18.6 years. The azimuths of the lunar stations always comprise those of the sun solstices, thus the minor lunar standstill has an azimuth which is always reached also by the sun, and the major lunar standstill has an azimuth which is never reached by the sun. It follows, that there is no way to distinguish an alignment to the minor lunar standstill from a solar alignment, while azimuths not belonging to the sun range may lead to major lunar standstills. However, the major northern standstill of the moon at the latitude of the two "exceptional" towns oriented at 45 degrees NE is around 41 degrees, and thus no hint of lunar orientation appears to be present.

## 4. Conclusion and perspectives

The results of the present paper show that the orientation of Roman towns in Italy is not random: it comprises two "families", one lying in the sector within ten degrees SE, the other near the winter solstice sunrise. Orientation of some towns to the sunrise in dates corresponding to important festivities of the Roman calendar, in particular *Terminalia*, looks also probable.

The existence of astronomical orientations confirms statements made by many Roman writers themselves, and raises the problem of the symbolic meaning of the *castrum* layout. It is indeed worth noticing that the idea of a quadripartite world reflected in a quadripartite human space occurred in many cultures, for instance among the Inca (the Inca state was just "the state of the four parts") or the Assires; in the italic context it was merged with astronomical orientation, perhaps by the Etruscans: Misa is indeed astronomically oriented to the cardinal points with a good precision, while cardinal orientation is barely visible in Greek towns.

Of course, this problem is open to future investigations. However, I would like to mention two examples which point out the symbolic relevance of the "squared" mentality among the Romans. The first example is the famous "magic square" graffito which was found in the Pompeii *Palestra*, dated to the period 63-79 a.D. and composed by a palindrome of five words of five letters (*ROTAS OPERA TENET AREPO SATOR*). I will not, of course, enter here in the important and unsolved problem if the graffito is an early testimony of the Christian religion (an interpretation which relies on the fact that the words can be rearranged to give a *cross* composed by the words PATERNOSTER together with an A and a O); what I would like to stress is only that the graffito *in any case* seems to be a symbol connected with the foundation ritual and the structure of the castrum. Indeed, the word *TENET* forms the image of a castrum, and the argument of the inscription refers to a sower (a *SATOR*, perhaps called *AREPO*) who is probably making some - perhaps ritual – ploughing (*ROTAS OPERA TENET*).

The second example is connected with the symbolic content of the *castrum* in its proper meaning, namely the military camp. It is the layout of the Trajan forum, constructed around 112 a.D.: the architect indeed conceived it as a replica of castrum military camp (see Fig. 3). Recent investigations have been carried out on the orientation of the Roman camps and forts (Richardson 2005) and, although precise indications are also in this case difficult to be obtained due to the relative low number of data sample (see Peterson, 2007, and Salt, 2007) there is a clear tendency to the orientation to the cardinal points which is hardly justifiable with strategic reasons.

All in all, in the author's view it is highly desirable that future research may dispose of a complete database of orientations – including temples - in the Roman world, a huge task which however would be of fundamental help towards a re-assessment of the (usually neglected) astronomical knowledge among the Romans, as well as to a better understanding of his connection with the Roman religion and mentality.

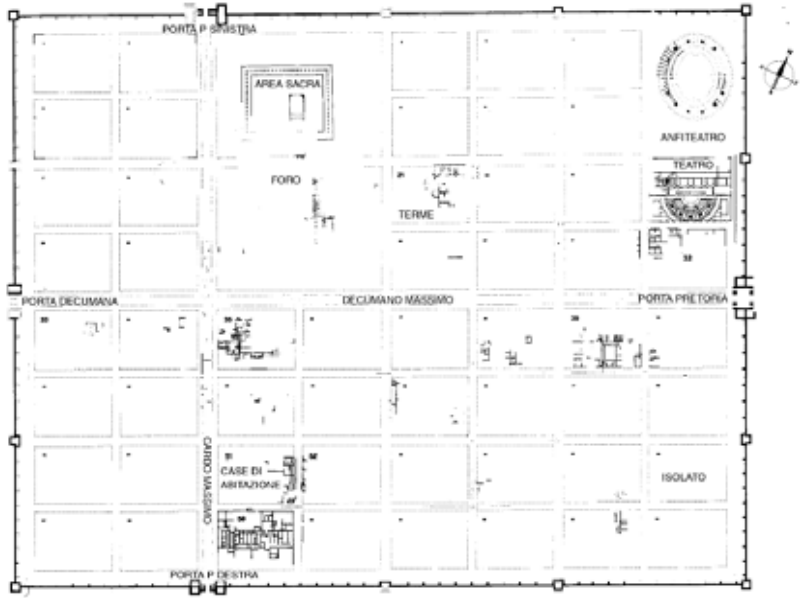

Fig. 1 Plan of Augusta Pretoria (Aosta)

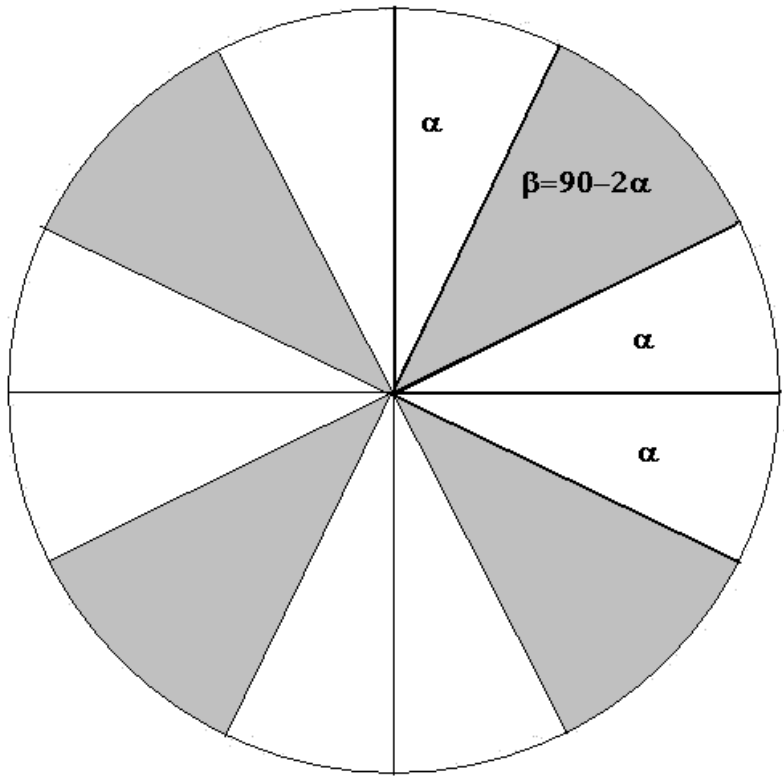

Fig. 2. The range of the rising positions of the sun at a certain latitude, spanned by the angle α, and the sector of possible non-solar orientations of the axes of a castrum town at the same latitude, spanned by the angle β=90-2α. See Sec. 3 for details.

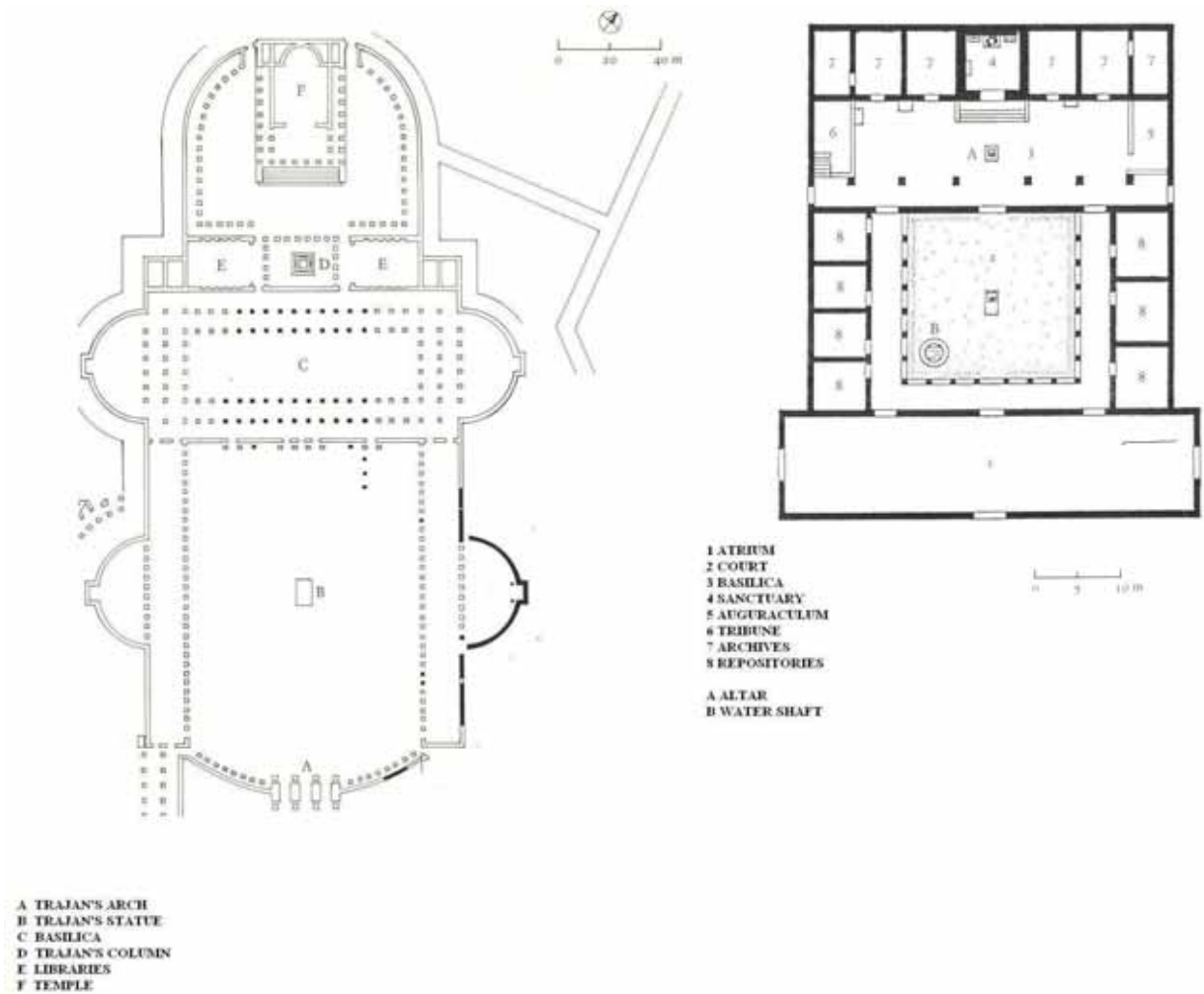

Fig. 3 Plan of Trajan's Forum (left) compared with a Roman military camp (right) (adapted from Settis et al 1988).

Table 1. Orientation of 38 Roman towns in Italy.

| Lat | Name | Location | Orientation | | |
|---|---|---|---|---|---|
| 43.55 | PISAURUM | Pesaro (184 b.C.) | 44 NE | 33 | |
| 43.42 | SENA GALLICA | Senigallia (284 b.C.) | 44 NE | 33 | |
| 44.03 | ARIMIMIUM | Rimini (268 b.C.) | 37 NE | 34 | |
| 45.26 | VERONA | Verona, pre-roman. Roman plan I century b.C. | 36 NE | 35 | Y |
| 45.33 | VICETIA | Vicenza, pre-roman, roman plan around 135 b.C. | 34 NE | 35 | Y |
| 41.29 | VENAFRUM | Venafro, pre-roman, roman plan II century b.C. | 28 NE | 32 | |
| 45.44 | AUG.PRETORIA | Aosta (25 b.C.) | 26 NE | 35 | |
| 45.46 | AQUILEIA | Aquileia (II century b.C.) | 19 NE | | |
| 41.13 | TELESIA | Telese (near Benevento) | 19 NE | | |
| 40.38 | SURRENTUM | Sorrento, greek, roman plan I century b.C. | 17 NE | | |
| 41.44 | OSTIA | Ostia (IV century b.C., data from the first layout) | 17 NE | | |
| 41.16 | FORMIAE | Formia (III century b.C.) | 17 NE | | |
| 44.48 | PARMA | Parma (183 b.C.) | 8 NE | | |
| 46.04 | TRIDENTUM | Trento (23 b.C.) | 8 NE | | |
| 41.44 | ALETRIUM | Alatri, pre-roman, Roman plan around II century b.C. | 7 NE | | |
| 42.35 | HATRIA | Atri (near Teramo) pre-roman, colony in 289 b.C. | 7 NE | | |
| 40.26 | PAESTUM | Near Salerno; greek, colony in 273 b.C. | 2 NE | | |
| 43.47 | FLORENTIA | Firenze (around 59 b.C.) | 1 SE | | Y |
| 41.06 | SINUESSA | Mondragone (296 b.C.) | 3 SE | | |
| 43.51 | LUCA | Lucca (around 180 b.C.) | 4 SE | | |
| 40.49 | PUTEOLI | Pozzuoli - Rione Terra (194 b.C.) | 4 SE | | |
| 42.18 | FALERII NOVI | Near Civita Castellana (240 b.C.) | 4 SE | | |
| 42.50 | ASCULUM | Ascoli Piceno, pre-roman, Roman plan 269 b.C. | 6 SE | | |
| 43.47 | ALBINTIMILIUM | Near Ventimiglia (I century b.C.) | 7 SE | | |
| 41.50 | PRAENESTE | Palestrina, pre-roman, Roman plan around 90 b.C. | 7 SE | | |
| 45.32 | BRIXIA | Brescia, pre-roman, Roman plan I century b.C. | 8 SE | | |
| 44.30 | BONOMIA | Bologna, pre-roman, Roman colony 189 b.C. | 12 SE | | Y |
| 41.41 | FERENTINUM | Ferentino, pre-roman, Roman plan II century b.C. | 17 SE | | |
| 42.17 | PELTUINUM | Near Prata D'Ansidonia (I century b.C.) | 18 SE | | |
| 45.04 | AUG.TAURINORUM | Torino (29 b.C.) | 30 SE | 34 | |
| 41.16 | MINTURNAE | Minturno (296 b.C.) | 31 SE | 32 | |
| 41.35 | NORBA | Near Norma (IV century b.C., layout II century b.C.) | 32 SE | 32 | Y |
| 44.33 | AUG.BAGIENNORUM | Bene Vagienna (I century b.C.) | 36 SE | 34 | Y |
| 42.06 | ALBA FUCENS | Massa d'Albe (IV century b.C.) | 37 SE | 32 | |
| 40.17 | GRUMENTUM | Grumento Nova (III century b.C.) | 37 SE | 31 | |
| 42.25 | COSA | Ansedonia (III century b.C.) | 38 SE | 33 | |
| 44.05 | LUNI | Luni (177 b.C.) | 38 SE | 34 | |
| 43.50 | FANUM | Fano, pre-roman, Roman plan I century b.C. | 39 SE | 33 | |